\documentclass[12pt,twoside,a4paper,fleqn]{article}
\usepackage[left=25mm,top=20mm,right=14mm,bottom=25mm]{geometry}
\usepackage{epsfig}
\usepackage{graphicx}
\usepackage{amssymb}
\usepackage{mathrsfs}
\usepackage{dcolumn}
\usepackage{amsmath}
\usepackage[numbers,sort&compress]{natbib}
\usepackage[utf8]{inputenc}
\usepackage[T1]{fontenc}
\usepackage{setspace}
\usepackage{multirow}
\usepackage{hyperref}
\usepackage{appendix}
\usepackage{float}
\DeclareMathOperator\erf{erf}
\newcommand{\pr}{\partial}

\newcommand{\rta}{\rightarrow}

\newcommand{\ep}{\epsilon}

\newcommand{\beq}{\begin{equation}}
	\newcommand{\eeq}{\end{equation}}

\newcommand{\ball}{\begin{align}}
	\newcommand{\eall}{\end{align}}

\newcommand{\beqar}{\begin{eqnarray}}
	\newcommand{\eeqar}{\end{eqnarray}}

\newcommand{\ben}{\begin{enumerate}}
	\newcommand{\een}{\end{enumerate}}

\begin{document}
	\date{}
	\title{Density-Driven Resistance Response in $MnS_{2}$: Theory} 
	\author{%
		{Komal Kumari$^*$, Raman Sharma$^*$ and Navinder Singh$^{**}$}\\
		$^*$\small Department~of~Physics,~\small Himachal~Pradesh~University,\\~\small Shimla,~India, Pin:171005.\\$^{**}$ \small Physical Research Laboratory, Ahmedabad,\\~\small Gujarat, India, Pin: 380009.\\
		 {\small komal.phyhpu@gmail.com;~raman.sharma@hpuniv.ac.in;~navinder@prl.res.in} 
	}		
\maketitle

	\begin{abstract}
		A colossal insulator-to-metal transition  in high-spin pyrite phase of $MnS_{2}$ has been experimentally observed \cite{colomns2}.  There are two possibilities behind this colossal insulator-to-metal transition: (1) migration of $Mn$ electrons to unoccupied $S^{2-}_{2}$ antibonding states under pressure which leads to conducting ligand states  and hence metallic transition, and (2) possibility of band crossing transition. We have analyzed this experimental obervation theoretically using a toy statistical model and found that the transition is  due to the migration of electrons from the transition metal ions to the ligand sites (i.e. the possibility (1)). The calculated resistivity compares well with the experimental data within the fitting parameters of the model.
	\end{abstract}
	
	\section{Introduction}
	\setstretch{1.25}
	Pressure-induced insulator-to-metal transition has been experimentally observed in pyrite-structured mineral $MnS_{2}$\cite{colomns2,gaintmns2,durkee2019}. $MnS_{2}$ is  a high spin ($S_{Mn}=\frac{5}{2}$) transition metal chalcogenides insulator, which undergoes to metallic state following a colossal drop in resistivity (order $\sim 10^{8}\Omega $) under pressure($\simeq 12GPa$). 
	At very high pressure ($P\gtrsim 36 GPa$) it is a low spin ($S_{Mn}=\frac{1}{2}$)  arseno-pyrite ($a-MnS_{2}$). The $a-MnS_{2}$ shows insulation type resistive behaviour ($\rho(T)$) increases as tempertaure decreases. \\
	
	In  ref. \cite{colomns2} authors  have proposed two mechanisms behind the observed transition. First one is that  the metallic state arises when unoccupied  disulfide $S^{-2}_{2}$ $\sigma^{*}_{3p}$ antibonding states becomes partly filled due to migration of $d$ electrons from $Mn$ to $S$. The second possibility is due to band crossing transition wherein occupied ligand $p$ band merges with the unoccupied metal $d$ band \cite{Q.Y., Kristin}. It helps the conduction band to migrate below the Fermi level near the $\Gamma$ point. This leads to the conducting behaviour of the system.\\
	 	
Under very high pressure up to $\sim 36GPa$, $MnS_{2}$ reveals a low spin state transition into the dense arsenopyrite phase accompanied by a giant volume collapse \cite{CHATTOPADHYAY1986305,Timirgazin_2016,Bros,rohr,allen}. This high density phase promotes electron localization effect and breaks the conduction. This leads to again high resistivity at higher pressure. \\
	
We purpose a statistical model to capture this insulator-to-metal transition \cite{loc-iti,kubo}. We consider localized electrons in $Mn$ $d$ orbitals which migrate to conducting ligands $S^{2-}_{2}$ sites under pressure. In the next section we present the mathematical formulation of our model.

	\section{Mathematical Formulation}
\begin{figure}{}
	\centering
	\includegraphics[height = 6.5cm, width =10cm]{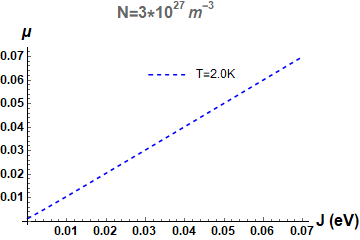}
	\caption{ Presents chemical potential as a function of $J$ at $T=2.0K$ }	\label{fig1}\end{figure}

	We consider the $N$ number of sites per unit volume on a three dimensional lattice \cite{loc-iti}. We assume single unpaired electron on each lattice site (this corresponds to localized electrons in $Mn$ $d$ orbitals). Let $J$ is the amount of energy cost to localize a given electron at a given site. If $n$ out of total $N$ electrons are in localized states (and $N-n$ will be in the itinerant ligand sites), then $n$ electrons cost $J n$ amount of energy to remain in localized sites. We write the canonical partition function for localized electrons as
	\beq
	\mathbb{Z}_{n}= \frac{N !}{n!(N-n)!} e^{-\beta J n}. \label{eq1}
	\eeq
	Here $\beta=\frac{1}{k_{B}T}$ is the inverse temperature.
	The  grand partition function for the system is
	\beqar
	\mathbb{Q}=\sum_{n=0}^{N}\mathbb{Z}_{n} e^{\beta \mu n}, \label{eq2}
	\eeqar
	where $\mu$ is the chemical potential. On substituting $\mathbb{Z}_{n}$ from (\ref{eq1}) in (\ref{eq2}), the grand partition function becomes
		\beqar
	\mathbb{Q}= \sum_{n=0}^{N}\frac{N !}{n!(N-n)!} e^{\beta n (\mu-J) }= (1+e^{\beta  (\mu-J)})^{N}. \label{eq3}
	\eeqar
Now the average number of the localized electrons is given by
\beq
 N_{Loc}=\frac{1}{\beta}\frac{\pr}{ \pr \mu}\log(	\mathbb{Q})= N\frac{e^{\beta  (\mu-J)}}{(1+e^{\beta  (\mu-J)})}. \label{eq4}
\eeq
The number of itinerant electrons can be written in the form $ N_{iti}=N- N_{loc}$:
\beq
 N_{iti} = N- N\frac{e^{\beta  (\mu-J)}}{(1+e^{\beta  (\mu-J)})}= \frac{N}{(1+e^{\beta  (\mu-J)})}. \label{eq5}
\eeq
These itinerant electrons form a Fermi sphere. Therefore in $3D$: 
\beq
N_{iti}(T)= \frac{2}{V} \sum_{k}f_{k}= \frac{2}{(2\pi)^3}\int \frac{d^3 k}{e^{\beta  (\ep-\mu)}+1}=\frac{8\pi}{(2\pi)^3}\int_{0}^{\infty} \frac{k^2 d k}{e^{\beta  (\ep-\mu)}+1}, \label{eq6}
\eeq
replacing $k$ integral with energy $\ep$ ($k^2=\frac{2 m \ep}{\hbar^2})$, we obtain
\beq
N_{iti}(T)= \frac{1}{2\pi^2}\frac{(2m)^{\frac{3}{2}}}{\hbar^3} \int_{0}^{\infty} \frac{\sqrt{\ep}d\ep}{e^{\beta(\ep-\mu)}+1}. \label{eq7}
\eeq	
The energy integral can be computed as 
\beq
N_{iti}(T)=\frac{1}{2\pi^2}\frac{(2m)^{\frac{3}{2}}}{\hbar^3}\bigg\{ \int_{0}^{\mu} \sqrt{\ep}d\ep+\int_{\mu}^{\infty} \frac{\sqrt{\ep}d\ep}{e^{\beta(\ep-\mu)}}\bigg\}. \label{eq8}
\eeq	
Under the relevant  low temperature condition $ T\rta 0$, $\beta\rta\infty$ it simplifies to
\beq
N_{iti}(T)=\frac{1}{2\pi^2}\frac{(2m)^{\frac{3}{2}}}{\hbar^3}\bigg\{\frac{2}{3}\mu^{\frac{3}{2}}+\frac{\sqrt{\mu}}{\beta}+\frac{\sqrt{\pi} e^{\beta \mu }}{2\beta^{\frac{3}{2}}}\erf(\sqrt{\beta \mu}) \bigg\}. \label{eq9}
\eeq
Now, the chemical potential can be obtained from the above equation and using equation (\ref{eq5}) for $N_{iti}(T)$:
\beq
\frac{N}{(1+e^{\beta  (\mu-J)})}=\frac{1}{2\pi^2}\frac{(2m)^{\frac{3}{2}}}{\hbar^3}\bigg\{\frac{2}{3}\mu^{\frac{3}{2}}+\frac{\sqrt{\mu}}{\beta}+\frac{\sqrt{\pi} e^{\beta \mu }}{2\beta^{\frac{3}{2}}}\erf(\sqrt{\beta \mu}) \bigg\} \label{eq10}
\eeq
The above expression can be computed  numerically to find $\mu$ for given values of T, J, and N. Fig \ref{fig1} shows $\mu$ as a function of $J$. 

\floatplacement{figure}{H}


Our aim is to study the dependence of resistivity ($\rho$) on $J$. We assume that the resistivity is given by the Drude-Lorentz model $\rho(J)=\frac{m}{N_{iti}(J) e^2 }\frac{1}{\tau}$. As $N_{iti}(J)$ has dependence on $J$, the change in $J$ leads to change in resistivity. The local repulsion parameter $J$ is pressure dependent. Under pressure there is a volume collapse and bond lengths (lattice paramenters) decrease. This shorter lattice parameters leads to more local repulsion thus larger $J$. In general there could be a complex dependence of $J$ on pressure i.e. $P=f(J)$.  In our model we consider a linear approximation and set $P=\alpha J$, where $\alpha$ is a constant to be determined by comparing the model with the experiment. We give the following justification for this linear dependence. Experiments \cite{colomns2} show that under maximum  pressure of $36GPa$ the volume reduces by $20\%$ and hence the lattice parameter is  reduced roughly by $6\%$. Now the dependence of $J$ on lattice parameter ($a$) can be roughly represented by  $J(\delta a)\sim \frac{1}{4\pi \varepsilon_{0} \varepsilon_{r}}(\frac{q_{1} q_{2}}{a+\delta a})$ $\sim k_{1} \frac{\delta a}{a}+ k_{2}$, where $ k_{1}$  and $ k_{2}$ are constant and $q_{1}$ and $q_{2}$ are charges on adjacent ions. Thus  in the leading order approximation we can set $J\sim \delta a $. Therefore under a very small change ($\delta a$) in the lattice parameter, the dependence of $J$  on $a$ can be taken as linear. This motivates our assumption $P=\alpha J$. With this assumption we compute $\rho$ and compares it with the experimental data. 
	\section{Experimental Comparison}
\begin{figure}{}
		\centering
		\includegraphics[height = 6.5cm, width =10cm]{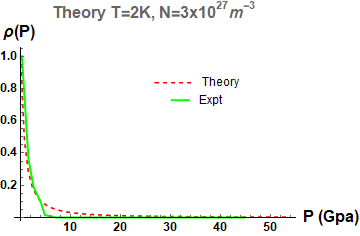}
		\caption{ Comparison of the present theory with the experimental data of  $MnS_{2}$}
		\label{fig3:j}
	\end{figure}

	Experimental data is obtained by digitizing the data in fig 1(a) of reference \cite{colomns2}. We normalize the experimental data as  $\frac{\rho(P)}{\rho(0)}$ as we are interested in the pressure evolution of resistivity not absolute magnitude. The data is shown with the solid green line in figure \ref{fig3:j}
	\floatplacement{figure}{H}
We  compute the resistivity from the formula $\rho(J)=\frac{m}{N_{iti}(J) e^2 }\frac{1}{\tau}$ where $P=\alpha J$. By treating $\alpha$ as our fitting parameter we plot $\rho(P)$ in figure \ref{fig3:j} (dashed red line). We use $\tau=10femtosec$ (a typical value of scattering rate in metals). Best fitting of our model with data corresponds to $\alpha=76.92$ $GPa/eV$. This leads to the validation of our model within the above mentioned assumption. 
 \floatplacement{figure}{H}



\section{Conclusion}	
Our simple statistical toy model which describes the localized electron to itinerant electron transition with the change of local repulsion parameter can qualitatively explain the mechanism of colossal resistive transition in $MnS_{2}$. Our model calculation shows that it is the migartion of $Mn$ electrons to unocuupied $S^{2-}_{2}$ antibonding states under pressure which leads to conducting ligand states and hence the metallic transition.

	\end{document}